\documentclass[sigconf]{acmart}
\usepackage{dsfont}
\usepackage{adjustbox}
\usepackage[utf8]{inputenc}
\usepackage{amsmath,amssymb}  
\usepackage[mathscr]{eucal}
\AtBeginDocument{%
  \providecommand\BibTeX{{%
    \normalfont B\kern-0.5em{\scshape i\kern-0.25em b}\kern-0.8em\TeX}}}

\setcopyright{rightsretained}
\acmConference[SIGIR eCom'20]{ACM SIGIR Workshop on eCommerce}{July 30, 2020}{Virtual Event, China}
\acmYear{2020}
\copyrightyear{2020}
\makeatletter
\renewcommand\@formatdoi[1]{\ignorespaces}
\makeatother
\acmISBN{}
\begin{document}

\title[Fantastic Embeddings and How to Align Them]{Fantastic Embeddings and How to Align Them: Zero-Shot Inference in a Multi-Shop Scenario}

\author{Federico Bianchi}
\authornote{Main authors, contributed equally to ideation and execution of this research project and are listed alphabetically.}
\affiliation{%
  \institution{Bocconi University}
  \streetaddress{Via Roberto Sarfatti, 25}
  \city{Milano}
  \country{Italy}}
\email{f.bianchi@unibocconi.it }

\author{Jacopo Tagliabue}
\authornotemark[1]
\authornote{Corresponding author.}
\affiliation{%
  \institution{Coveo Labs}
  \city{New York}
  \state{NY}
  \postcode{10002}
}
\email{jtagliabue@coveo.com}

\author{Bingqing Yu}
\authornotemark[1]
\affiliation{%
  \institution{Coveo}
  \city{Montreal}
  \country{Canada}}
\email{cyu2@coveo.com}

\author{Luca Bigon}
\authornote{Author was responsible for data ingestion and data engineering.}
\affiliation{%
  \institution{Coveo}
  \city{Montreal}
  \country{Canada}}
\email{lbigon@coveo.com}

\author{Ciro Greco}
\authornote{Author was responsible for analysis of the
downstream NLP task.}
\affiliation{%
  \institution{Coveo Labs}
  \city{New York}
  \state{NY}
  \postcode{10002}
 }
\email{cgreco@coveo.com}

\renewcommand{\shortauthors}{Bianchi et al.}

\begin{abstract}
  This paper addresses the challenge of leveraging multiple embedding spaces for multi-shop personalization, proving that zero-shot inference is possible by transferring shopping intent from one website to another without manual intervention. We detail a machine learning pipeline to train and optimize embeddings \textit{within shops} first, and support the quantitative findings with additional qualitative insights. We then turn to the harder task of using learned embeddings \textit{across shops}: if products from different shops live in the same vector space, user intent - as represented by regions in this space - can then be transferred in a zero-shot fashion across websites. We propose and benchmark unsupervised and supervised methods to ``travel'' between embedding spaces, each with its own assumptions on data quantity and quality. We show that zero-shot personalization is indeed possible at scale by testing the shared embedding space with two downstream tasks, event prediction and type-ahead suggestions. Finally, we curate a cross-shop anonymized embeddings dataset to foster an inclusive discussion of this important business scenario.
\end{abstract}



\maketitle

\section{Introduction}
Inspired by the similarity between words in sentences and products in browsing sessions, recent work in recommender systems re-adapted the NLP CBOW model \cite{Mikolov2013EfficientEO} to create \textit{product embeddings}~\cite{Lake19}, i.e. low-dimensional representations which can be used alone or fed to downstream neural architectures for other machine learning tasks. Product embeddings have been mostly investigated as static entities so far, but, exactly as words \cite{Bianch2019}, products are all but static. Since the creation of embeddings is a stochastic process, training embeddings for similar products in different digital shops will produce embedding spaces which are not immediately comparable: how can we build a unified cross-shop representation of products? In \textit{this} work, we present an end-to-end machine learning pipeline to solve the transfer learning challenge in digital commerce, together with substantial evidence that the proposed methods -- even with no supervision -- solve effectively industry problems that are otherwise hard to tackle in a principled way (e.g. zero-shot inference in a multi-shop scenario).

We summarize the main contributions of \textit{this} paper as follows:
\begin{itemize}
\item we extensively investigate product embeddings in both the \textit{within-shop} and \textit{cross-shop} scenarios. Since this is the \textit{first} research work to tackle \textit{cross-shop} inference by aligning embedding spaces, Section~\ref{usecasesection} explains the use cases at length. While \textit{within-shop} training is not a novel topic \textit{per se} (Section~\ref{relatedwork}), we report detailed quantitative results since we could \textit{not} replicate previous findings in hyperparameter tuning; we also improve upon existing pipelines by proposing a qualitative validation as well;
\item we propose, implement and benchmark several aligning methods, varying the degree of supervision and data quality required. We provide quantitative and qualitative validation of the proposed methods for two downstream tasks: a ``next event prediction'' and a type-ahead personalization task, in which aligned embeddings are used as input to a conditional language model;
\item curate and release in the public domain a cross-shop product embeddings dataset\footnote{At the time of drafting \textit{this} paper, discussions within the legal team of \textit{Coveo} are still ongoing to settle on a final license for the data; as such, dataset details may change before final publication: feel free to reach out to us for any update.} to foster reproducible research on this topic. With practitioners in the industry in mind, we also detail our cloud architecture in Appendix~\ref{paasdetailssection}.
\end{itemize}
Our analysis of product data from several stores found that product embeddings, while superficially similar to word embeddings, have their own peculiarities, and data assumptions need to be assessed on a case-by-case basis. Moreover, our benchmarks confirm that the proposed methodology is of great interest when a single SaaS provider can leverage cross-client data, or when a multi-brand/multi-regional group can use data from one store to improve performance on another.

\section{Use cases from the industry}
\label{usecasesection}
\begin{figure}
  \centering
  \includegraphics[width=\linewidth]{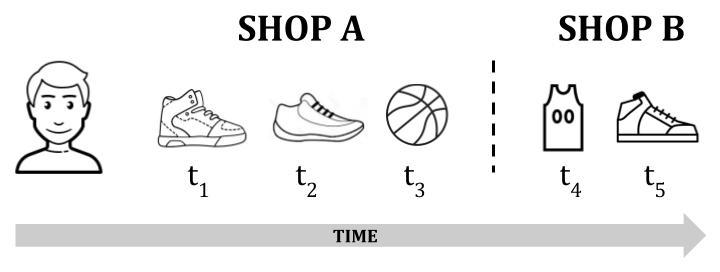}
  \caption{Cross-shop use case: a user browsing "basketball-related" products on \textbf{Shop A} and then continuing the session on \textbf{Shop B} with similar products.}
  \label{use_pic}
  \Description{Cross-shop use case.}
\end{figure}

Shoppers are likely to browse in multiple related digital shops before making the final purchase decision, as most online shopping sessions (as high as 99\% \cite{Gudigantala2016}) do not end with a transaction. The cross-shop scenario depicted in Fig.~\ref{use_pic} is therefore very common: the shopper starts browsing on \textbf{Shop A} for basketball products and ends up continuing his session on \textbf{Shop B}. 

Providing relevant content to unknown shoppers is of paramount importance to increase the probability of a conversion, considering that e-commerce websites tend to have high bounce rates (i.e. average percentage of users who leave after a single interaction with the page ranges between 25\% and 40\% ~\cite{SimilarWeb2019}) and low ratios of recurring customers (<9\% in our dataset). Moreover, there is vast consensus in the industry on the importance of personalization \cite{SCHREINER201987} in boosting the quality of the shopping experience and increasing revenues: but how is it possible to personalize the experience of a user that has never been on the target site?

The rationale for \textit{this} research work is thus the importance of providing personalized experiences \textit{as early as possible} and with \textit{as little user data as possible}: generally speaking, we propose to leverage the aligned product embedding space to model shopper's intent during a session -- if cross-shop browsing is, so to speak, a walk through the (aligned) product space, we can feed users's position to downstream neural systems to capture their shopping intent.

\begin{table}
  \caption{A sample of multi-brand retailers from Fortune 500.}
  \label{tab:fortune500}
  \begin{tabular}{lccc}
    \toprule
    Group & Rev. (M\$) & Brands & Examples\\
    \midrule
    TJX  & 41.717 & 7 & HomeSense, Marshalls\\
    Nike & 39.117 & 4 &  Converse, Nike\\
    Gap & 16.383 & 9 & Gap, Old Navy\\
    VF & 13.870 & 19 & Eastpack, Napapijri\\
    L Brands & 12.914 & 3 & Victoria Secret, Pink\\
    Hanesbrands & 6.966 & 29 & Champion, Playtex\\
  \bottomrule
\end{tabular}
\end{table}

There are two types of players which would naturally benefit from cross-shop personalization. The first is retail groups who own and operate multiple brands and shops (e.g. Gap Inc owns and operates Gap, Old Navy, etc). To give an idea of the size of this market share, the combined revenues generated by \textit{Fortune 500} retail groups with these characteristics is more than 130 billion dollars (see Table~\ref{tab:fortune500}). For these retailers, a portion of the user base consistently shops across different websites of the same group and it would be therefore beneficial to them to implement optimization strategies across multiple websites. Given the size of the market, it is easy to see how the implementation of successful personalization strategies across shops would translate into remarkable business value. At the same time, most of these groups are ``traditional'' retailers (as opposed to digitally native companies e.g. Amazon). Therefore, even if they would be benefiting the most from a unified view of their customers across different digital properties, in practice they are more likely to experience roadblocks related to technology. To this extent, the immediate value of the present work is to show for the first time that personalization across shops can be achieved even with minimal data tracking, no meta-data and no human intervention. The traditional nature of these retailers may also explain why cross-shop behavior is a niche use case in the research community, whose agenda is mostly set by tech companies – by publishing our findings we wish the community would join us in tackling this important use case. 

The second type of players are multi-tenant SaaS providers who provide AI-based services. For these companies the main challenge is to scale quickly within the verticals and minimize the friction in deployment cycles: being able to leverage some kind of “network effect” to transfer knowledge from one client to another would certainly be a distinctive competitive advantage. Recently, AI SaaS providers for e-commerce have received great attention from venture capitalists. As an indication of the size of the market opportunity, only in 2019 and only in the space of AI-powered search and recommendations, we witnessed \textit{Algolia} raising USD110M \cite{AlgoliaRound}, \textit{Lucidworks} raising USD100M \cite{LWRound} and \textit{Coveo} raising CAD227M \cite{CoveoRound}. While a full cross-shop data strategy depends on many non-technical assumptions (see Section~\ref{dataset_section} for a discussion of legal constraints), it is important to realize that some multi-property retail groups turn to external providers for certain AI services. While our methods do not assume any common meta-data between target shops (e.g. the two shops can be even in \textit{different} language), we expect our models to work better with catalogs that have significant ``semantic overlap'' (e.g. two shops selling sport apparel, Section~\ref{dataset_section}).

\begin{figure}
  \centering
  \includegraphics[width=\linewidth]{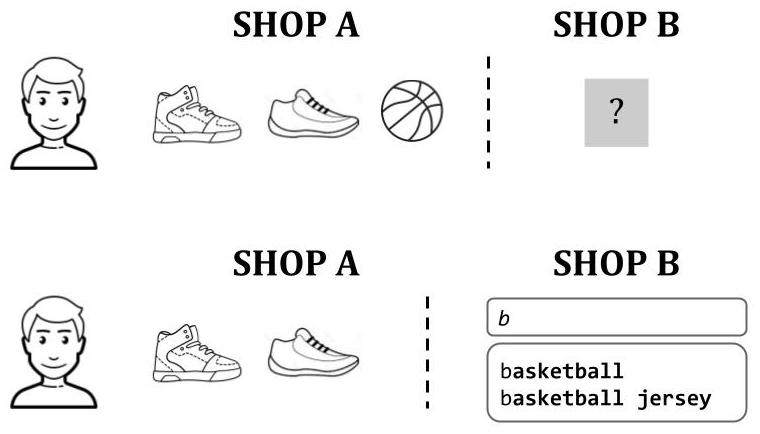}
  \caption{Zero-shot prediction tasks can be solved by transferring shopper's intent to the target website leveraging aligned embeddings; in the first task, the system predicts product interactions on \textbf{Shop B} from user browsing products on \textbf{Shop A}; in the second task, products on \textbf{Shop A} are used to personalize type-ahead suggestion on \textbf{Shop B} -- since the session is basketball-themed, we expect the system to promote (\textit{ceteris paribus}) basketball queries.}
  \label{pred_pic}
  \Description{Cross-shop prediction task.}
\end{figure}

We show several effective methods to achieve transfer learning across shops, each making different assumptions about data quantity and quality available. As discussed at length in Section~\ref{method_section}, a distinguishing feature of this use case is that we make no assumption \textit{at all} about catalog overlap (i.e. the shops involved can have 0 items in common), making it much more challenging that the typical (and well-studied) retargeting use case (i.e. a shopper sees ads on \textit{Site X} for the same product she was viewing on \textit{Site Y}). Our most interesting result is proving that even without any cross-shop data, personalization on the target shop can be achieved successfully in a pure zero-shot fashion. To showcase the possibilities opened up by cross-shop embeddings, we demonstrate the effectiveness of the aligned space tackling two prediction tasks, as depicted in Figure~\ref{pred_pic}: if the system observes the behavior of a user on \textbf{Shop A}, can it predict what she is going to browse/type on \textbf{Shop B}? As we shall see, the answer is ``yes'' for both use cases.

\section{Related work}
\label{relatedwork}
The work sits at the intersection of several research topics. 

\textbf{Product Embeddings}. Word2vec~\cite{Mikolov:2013} was introduced in 2013 as a neural method to generate vector representations of words based on co-occurrences; soon, the model was adapted to the product space, where it found immediate use in recommender systems~\cite{Grbovic15}.~\cite{Vasile16} introduced \textit{Meta-Prod2Vec}: given our focus on cross-shop learning, we decided to \textit{not} use product information as there is no guarantee that two shops will have comparable metadata. \cite{Caselles18} first studied the role of hyperparameters in recommendation quality: we extensively investigate hyperparameters as well, but we improve upon their validation procedure and add a qualitative evaluations of the embedding spaces.

\textbf{Aligning Embedding Spaces}. The problem of learning a mapping between spaces has been widely explored in NLP. In fact, Alignment is important for language translation \cite{conneau2017word,Artetxe18}, to study language change \cite{bamman2014distributed,hamilton2016diachronic,szymanski2017temporal,yao2018dynamic,Bianch2019}.
However, as explained in Section~\ref{method_section}, the availability of unequivocal pairs of matching items in two spaces (e.g. \textit{uno} and \textit{one} in language translation) make vector space alignment in NLP significantly different from our use case.  ~\cite{Bose2019MetaGraphFS} is a recent work on zero and few shots prediction in a recommender setting across multiple ``spaces'': their problem is phrased as a meta-learning task over graphs representing different cities, while our work is focused on behavioral-based embeddings and sessions across multiple spaces. Possibly because of the maturity of data ingestion required to rebuild session data and the difficulty in finding suitable datasets for experimentation, \textit{this} work is the first to our knowledge to extensively study product embeddings across multiple spaces.

\textbf{Deep Learning in Type-ahead Systems}. Suggest-as-you-type is a well studied problem in the IR community~\cite{Cai:2016:SQA:3099948}. Recent works have embraced neural networks:~\cite{inproceedingsparj2017} introduces a char-based language model,~\cite{conf/sigir/WangZMDK18} applies RNN to a noisy channel model (but the inner language model is not personalized like our proposed method). Specifically in e-commerce,~\cite{DBLP:journals/corr/abs-1905-01386} uses \textit{fastText} to embed previous queries and then re-ranks suggestions accordingly: our personalization layer does not require linguistic resources or previous queries, as the vast majority of sessions (>90\% in our network) for mid-size shops do \textit{not} contain search queries.~\cite{CoveoECNLP20} is the first exploration of cross-shop type-ahead systems, obtaining transfer learning by placing products in the same space through shared image features. The proposed~\textit{prod2vec} embeddings significantly outperform image-based representations to produce accurate conditional language models (18\% MRR improvement over the same shop).

\section{Dataset}
\label{dataset_section}
\textit{Coveo} is a Canadian SaaS provider of search and recommendation APIs with a global network of more than 500 customers, including several \textit{Fortune 500} companies. For \textit{this} research, we leverage behavioral data collected over 12 months from two mid-size shops (revenues >10M and <100M) in the same vertical (sport apparel); we refer to them as~\textbf{Shop A} and~\textbf{Shop B}. Data is sessionized by the pipeline after ingestion: \textit{prod2vec} embeddings are trained on product interactions that occur within each recorded shopper session (Section~\ref{prod2vec}). In the interest of practitioners in the industry, we share details on our cloud design choices in Appendix~\ref{paasdetailssection}.

Catalogs from \textbf{A} and \textbf{B} were also obtained to perform a qualitative check on our validation strategy and test semi-supervised approaches. After cleaning user sessions from bot-like behavior and sampling, descriptive statistics for the final product embedding dataset can be found in Table \ref{tab:descriptive}; even if \textbf{A} and \textbf{B} differ in catalog size and traffic, they have <9\% of \textit{recurring} customers (i.e. shoppers with more than 3 sessions in 12 months).  

\begin{table}
  \caption{Descriptive stats for \textbf{Shop A} and \textbf{Shop B}}
  \label{tab:descriptive}
  \begin{tabular}{lccc}
    \toprule
    Shop & Sessions (events) & SKUs & 25/50/75 pct\\
    \midrule
    A  & ~3M (10M) & ~23k & 3, 5, 7\\
    B &~11M (32M) & ~42k &  3, 5, 7\\
  \bottomrule
\end{tabular}
\end{table}

We believe it is important to explicitly address two potential legal concerns about the underlying dataset of \textit{this} research:

\begin{itemize}
    \item end-user~\textit{privacy}: data collected is fully anonymized, in line with GDPR adequacy; data tracking required to produce aligned embeddings is \textit{significantly less} than other standard e-commerce use cases (e.g. re-targeting);
    \item data \textit{ownership}: the possibility to use aggregate (embeddings-based) data across websites depends on case-by-case legal constraints and specific contractual clauses. Websites operated by the same group have generally no issue in sharing data to improve overall performance. On the other hand, websites operate by different companies may see each other as competitors. In our experience, the answer is not clear-cut: mid-size shops (like \textbf{A} and \textbf{B}) tend to be less protective and more focused on the upside of a system that is aware of industry trends; bigger players, on the other side, seem to be more defensive; interestingly, the latter are more likely to have multi-brand deployment, making the methods here developed still relevant for many use cases.   
\end{itemize}

Finally, a sample of browsing sessions for distinct users with cross-shop behavior was obtained to benchmark different methods on the downstream prediction tasks: it is worth remembering that several proposed methods for cross-shop inference (Section~\ref{method_section}) do \textit{not} rely on cross-shop data, which is used in the unsupervised and semi-supervised case as gold standard only. 

\section{Methods}
\label{method_section}
The cross-shop inference is built in two phases. First, the system learns the best embeddings for \textbf{A} and \textbf{B} \textit{separately}, second, it learns a mapping function from one space to the other, implicitly aligning the two embedding spaces and enabling cross-shop predictions. 

\subsection{Learning optimal product embeddings}
\label{prod2vec}
Product embeddings are trained using CBOW with negative sampling~\cite{Mikolov:2013,Mikolov2013EfficientEO}, by swapping the concept of words in a sentence with products in a browsing session; for completeness we report a standard formulation~\cite{DBLP:journals/corr/abs-1804-00306}. For each product $p\in\mathcal{P}$, its center-product embedding and context-product embedding are \textit{d}-dimensional vectors in $\mathds{R}$, $\mathcal{U}$[p] and $\mathcal{V}$[p]: embeddings are learned by solving the following optimization problem:

\begin{equation}
\footnotesize{
\max\limits_{\substack{\mathcal{U}:\mathcal{P}\to\mathds{R}^d \\
\mathcal{V}:\mathcal{P}\to\mathds{R}^d}}\sum\limits_{(p,c)\in\mathcal{D}^{+}}\log \sigma\left(\mathcal{U}[p]^{\top}\mathcal{V}[c]\right)+\sum\limits_{(p,c)\in\mathcal{D}^{-}}\log \sigma\left(-\mathcal{U}[p]^{\top}\mathcal{V}[c]\right)
}
\label{eq:formula1}
\end{equation}

where ${D}^{+}$/${D}^{-}$ are positive/negative pairs in $D$, and $\sigma(·)$ is the standard sigmoid function. Following the findings in \cite{Caselles18}, we performed extensive tuning on the most important hyperparameters (Table \ref{tab:hyperdescription}) and develop both quantitative and qualitative protocols to evaluate the quality of the produced embedding space.

\begin{table}
  \caption{Hyperparameters and their ranges.}
  \label{tab:hyperdescription}
  \begin{tabular}{ll}
    \toprule
    Gensim Parameter & Tested Values\\
    \midrule
    \textit{min\_count} & 2, 3, 5, 10, 15, 30\\
    \textit{window} & 2, 3, 5, 10, 15\\
    \textit{iter} & 5, 10, 20, 30, 50\\
    \textit{ns\_exponent} & -1.0, -0.5, 0.0, 0.75, 1.0\\
  \bottomrule
\end{tabular}
\end{table}

\subsubsection{Quantitative validation}
We focused on a \textit{Next Event Prediction} (NEP) task to evaluate quantitatively the quality of the embeddings: given a session \textit{s} made by events $e_1$, ... $e_n$, how well $e_1$, ... $e_{n-1}$ can predict $e_n$?

\begin{figure}
    \includegraphics[width=0.20\textwidth]{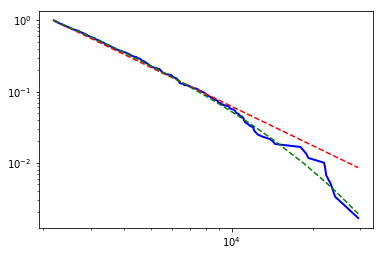}
    \includegraphics[width=0.20\textwidth]{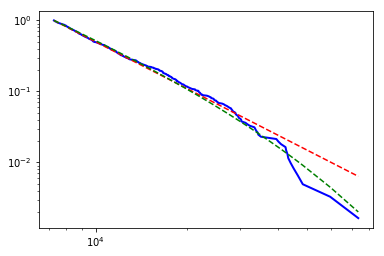}
    \caption{\textbf{Shop A} (left) and \textbf{Shop B} (right) log plots for product views: \textit{empirical} distribution is in blue, \textit{power-law} in red and \textit{truncated power-law} in green. Truncated power-law is a better fit than standard power-law for both shops ($p<.05$), with $\alpha=2.32$ for \textbf{A} and $\alpha=2.72$ for \textbf{B}. Power-law analysis and plots are made with \cite{Alstott_2014}.}.
    \label{pic:logdistributions}
\end{figure}

To address the NEP, we propose to use the entire session preceding the target event, by constructing a session vector averaging the embeddings for $e_1$, ... $e_{n-1}$ and then apply a \textit{Nearest Neighbors} classifier to predict $e_n$. Our choice is in contrast with what proposed by \cite{Caselles18}, which conducts hyperparameter tuning using kNN with just one item, $e_{n-1}$, as seed: from our experience in digital commerce, buying preferences are indeed multi-faceted, and important information about user intentions may be hidden at the start of the session (\cite{10.1145/2959100.2959190, CoveoECNLP20})\footnote{We also used LSTM as an alternative algorithm for validation, with similar results. We opted to report only kNN since a simpler model allows our valuation to be focused on the quality of the embeddings themselves, not so much the algorithm.}. Both \textbf{H@10} and \textbf{NDCG@10} were calculated for each trained model, but \textbf{NDCG@10} was primarily used for evaluation:

\begin{equation}
Discounted\ CG_k = DCG_k = \sum_{i=1}^{k}{\frac{rating(i)}{log_2(i + 1)}}
\label{eq:dcg}
\end{equation}

\begin{equation}
Ideal\ DCG_k = IDCG_k = \sum_{i=1}^{|REL|}{\frac{rating(i)}{log_2(i + 1)}}
\label{eq:idcg}
\end{equation}

\begin{equation}
NDCG_k=\frac{DCG_k}{IDCG_k}
\label{eq:ndcg}
\end{equation}

where ${|REL|}$ is the list of ground truth target events, up to ${k}$, and ${rating(i)}$ is the binary relevance value, which means ${rating(i) = 1}$ if event ${i}$ is found in the ground truth target events; otherwise, ${rating(i) = 0}$. Best and worst models, with parameters and score, can be found in Table \ref{tab:bestworstsettings}. It is interesting to remark that our extensive validation could \textit{not} confirm many generalizations put forward in \cite{Caselles18}: negative exponent was \textit{not} found to be a consistent factor in improving embeddings quality and \textbf{Shop A} and \textbf{Shop B} best parameter combinations are very similar, despite the underlying distribution being different (Figure \ref{pic:logdistributions}); moreover, the gap between best and worst models was found to be significant, but not as wide as \cite{Caselles18} indicated.

\begin{table}
  \caption{Best and worst parameter settings by shop, with validation score.}
  \label{tab:bestworstsettings}
  \begin{tabular}{lccccc}
    \toprule
    Model & Min Count & Window & Iter. & Exp. & NDCG@10\\
    \midrule
   \textbf{A - Best} & 15 & 10 & 30 & 0.75 & 0.1490\\
    \textbf{A - Worst} & 2 & 15 & 10 & -0.5 & 0.1058\\
   \textbf{B - Best} & 15 & 5 & 30 & 0.75 & 0.2452\\
   \textbf{B - Worst} & 5 & 10 & 30 & -0.5 & 0.1881\\
  \bottomrule
\end{tabular}
\end{table}

\subsubsection{Qualitative validation}
\label{qualitative_val}
The evaluation of word embedding models is intrinsically built on human-curated analogies such as \textit{boy} : \textit{king} = \textit{women} : \textit{?} \cite{pennington-etal-2014-glove} as both a quantitative check (``how many analogies can be solved by the vector algebra in the given space?'') and a qualitative one (``can we confirm, as humans, that the semantic properties captured by the space are indeed close to our linguistic intuitions?''). While analogies are indeed potentially meaningful in the product spaces for specific use cases (e.g. what is the Nike’s “air jordan shoes” equivalent for Adidas?), compiling a list for validation would be time-consuming and involving arbitrary choices.

\begin{figure}
    \includegraphics[width=0.225\textwidth]{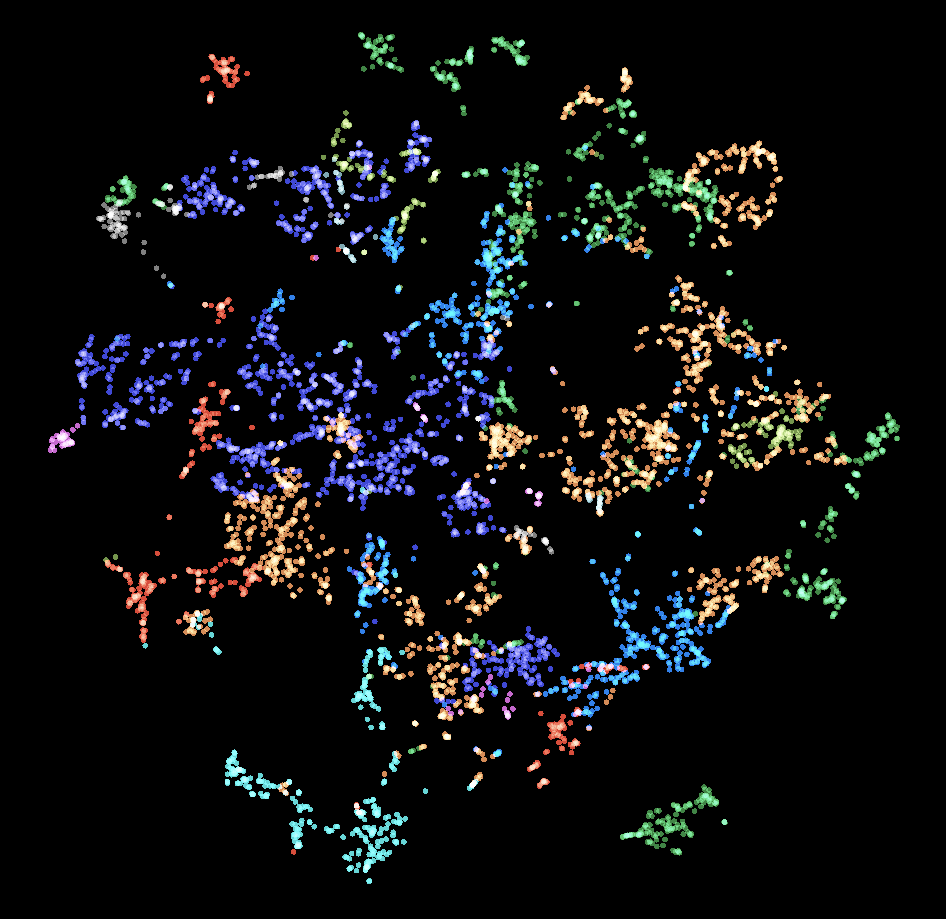}
    \includegraphics[width=0.22\textwidth]{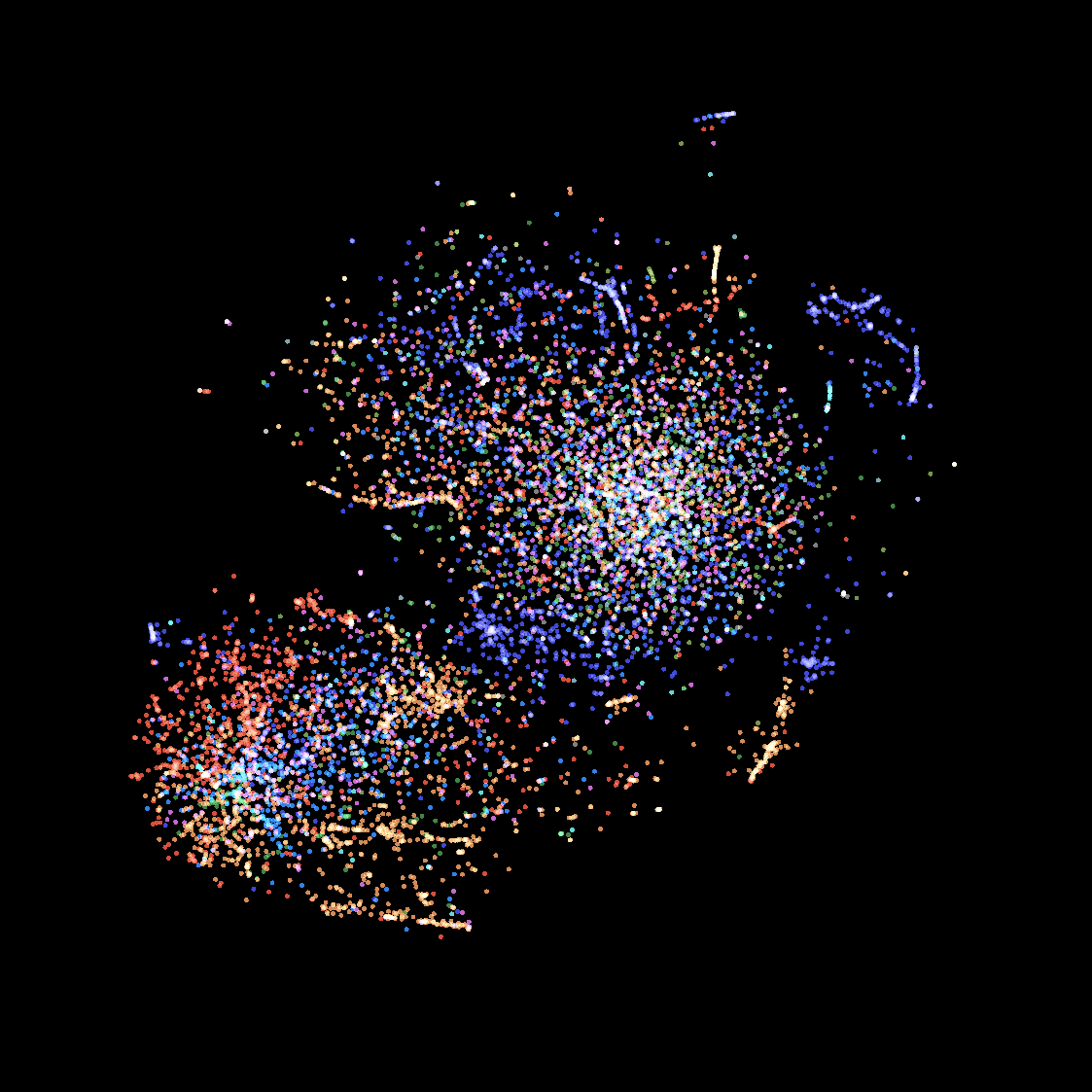}
    \caption{2-dimensional projections (t-sne) of high-scoring (left) and low-scoring (right) models according to the NEP task. Each point is a product in \textbf{Shop A} embedding space, color-coded by sport activity through catalog meta-data: it is easy to notice that high-scoring models produce sharper clusters in the embedding space. Projections are obtained with following parameters:~\textit{perplexity=25},~\textit{learning rate=10}, ~\textit{iterations=500}.}
    \label{pic:embeddingtsne}
\end{figure}

To have an independent qualitative confirmation that the NEP task is enforcing meaningful distinctions between spaces trained with different parameters, we sampled a model from the top 5 and one from bottom 5 in the NEP ranking, and leverage domain experts to classify products into sport activities (soccer, basketball, tennis, etc., for a total of \textit{N}=10 activities). We use t-sne~\cite{Maaten08visualizingdata} to project embeddings into two-dimensions and color-code the products with labels: as shown in Figure~\ref{pic:embeddingtsne}, better embeddings form sharper clusters with homogeneous coloring. To confirm the visual results, we train a Multilayer Perceptron (MLP) with the objective of predicting the activity from the embeddings\footnote{The MLP has two dense layers with \textit{relu} activation, a softmax layer for prediction, \textit{dropout} of $0.5$ between layers, \textit{SGD} as optimizer.}. Confirming the visual inspection, the accuracy score was $0.95$ for the high-performing model and $0.32$ for the low-performing one.

\subsection{Crossing the (shop) chasm}
\label{mapping_section}
Cross-embedding learning in the NLP space takes place in a continuum of supervision: from thousands of "true" pairs \cite{Mikolov2013ExploitingSA}, to dozen of them \cite{artetxe-etal-2017-learning}, to no pair at all \cite{conneau2017word}. However, it should be emphasized that aligning word spaces and aligning product spaces are \textit{not} the same task:

\begin{enumerate}
    \item given two languages, both will contain the same "semantic regions" (e.g., general topics like \textit{places}, \textit{animals}, \textit{numerals}, etc.) and, within those regions, several overlapping tokens (e.g. \textit{dog} is \textit{cane} in Italian, \textit{one} is \textit{uno}, \textit{lake} is \textit{lago}, etc.); however, given even shops in the same vertical such as \textbf{Shop A} and \textbf{Shop B}, there is no guarantee they will both contain products for, say, \textit{climbing};
    \item given two languages, there are linguistic resources mapping items from one to the other non-arbitrarily; however, given shops in the same verticals, finding exact duplicates is non-trivial and there are many cases in which mapping is arguably undetermined.
\end{enumerate}

It is also important to stress that \textit{no product is assumed to be the same across the two shops}: while we know \textbf{Shop A} and \textbf{Shop B} have comparable catalogs in terms of \textit{type} of items (e.g. they both sell sneakers, boots, etc.), we make no assumption about them having the same \textit{tokens} (i.e. we don't know if they both sell a specific pair of shoes, \textit{Air Zoom 95}), and we make no use of textual meta-data.\footnote{Assuming user and/or attribute overlap is the typical setting for cross-domain recommender systems~\cite{articleCrossDomain}: for this reason, they are not a meaningful baseline for the scope of \textit{this} work.}.

Considering those differences, we built and tested a wide range of unsupervised and supervised models to address the \textit{cross-shop} challenge:

\begin{itemize}
    \item \textit{image-based model} (IM), a completely unsupervised model using weak similarity signals derived from image vectors to build a "noisy" seed for a self-learning framework \cite{Artetxe18}. In particular, we sample images from \textbf{Shop A} and \textbf{Shop B} full catalogs and run through a pre-trained VGG-16 network \cite{Simonyan15} to extract features from the \textit{fc2} layer; PCA is then applied to reduce the feature dimensions from 4096 to $d$ dimensions; K-means is then used to group the vectors for \textbf{Shop A} into $k$ clusters: 2 points closest to the centroids of each cluster are the "sample points"; for each of these points, we use kNN to retrieve the closest image from \textbf{Shop B}. The seed dictionary built in this fashion is used to bootstrap the self-learning framework, and iteratively improve the mapping and the dictionary until convergence. It is worth noting that the alignment results reported below are achieved even if the seed dictionary is indeed noisy (as verified manually by sampling the quality of the pairings), witnessing the robustness of the proposed procedure. Different values for $d$ (5, 10, 20, 40, 60, 100) and $k$ (15, 30, 50, 70) were tested, but we report the scores for the best combination ($d=20$ and $k=50$). This method is both completely unsupervised \textit{and} fully "\textbf{zero}-shot" in the \textit{cross-shop} scenario, as no data on \textit{cross-shop} sessions is ever showed to the model during training;
    \item \textit{user-based model} (UM), a fully supervised model leveraging \textit{directly} users browsing on the two target shops. In particular, given the last product seen on the source shop and the first on the target shop, we learn to map the two products using linear regression that is then generalized to map all the embeddings of the source shop to the target shop.
    \item \textit{user-translation model} (TM), a fully supervised model that is using \textit{directly} shoppers browsing on the two target shops as if it was a bi-lingual parallel corpus. In particular, task is modelled after a different NLP architecture, sequence-to-sequence networks for machine translation \cite{Sutskever2014SequenceTS}: the intuition is that \textbf{Shop A} and \textbf{Shop B} behave quite literally as different languages and deep neural nets are well suited to learn how to encode in one space and decode in the other the latent intent of the shopper. We use the sequence to sequence model provided by the OpenNMT tool \cite{opennmt} that comes with 2-layer LSTM with 500 hidden units on both the encoder and the decoder layer. We initialize the embeddings of the layers using our product embeddings. The model is trained to translate the sequence of products seen in  \textbf{Shop A} into a sequence of product seen in \textbf{Shop B}.
\end{itemize}

By proposing and testing methods with different degrees of supervision, we provide substantial evidence that aligning embedding is possible in a variety of business scenarios: in particular, insofar as data tracking and technological capabilities vary across retailers, purely unsupervised methods (\textit{IM}) are particularly interesting as they make almost no assumption about available data and existing \textit{cross-shop} data points. On the other hand, supervised models (\textit{TM}) provide ``natural'' upper bounds for unsupervised counterparts, and can be deployed in business contexts where advanced data ingestion and data practices are already present. In general, our own experience is that these models can satisfy complementary business scenarios: for example, if historical fine-grained data is unavailable at day one (as it often is), aligning product embeddings with no cross-shop data is crucial to deliver personalization without advanced tracking capabilities. 

\section{Experiments}
\label{experiments}
We apply alignment methods to two downstream tasks: the first one is a straightforward extension to two shops of \textit{NEP}, as presented in Section~\ref{prod2vec} -- by aligning different product spaces, we hope to prove we can reliably guess shopper interactions with products on the target shop by transferring her intent from the first shop; the second task is an NLP-related task, in which aligned embeddings are used to build a conditional language model that can provide personalized suggestions to shoppers arriving at the target site~\cite{tagliabue-etal-2020-grow}: the query suggestion task is useful both to establish that \textit{prod2vec} transfer learning is superior to the image-based one~\cite{CoveoECNLP20}, and to prove that intent vectors are not just useful for recommendations, but also for a variety of personalization tasks in NLP.

It is important to highlight that our focus is to establish for the \textit{first} time that aligning product embeddings allow to transfer shopper intent between shops in scalable and effective ways; for this reason, we picked architectures which are straightforward to understand, in order to make sure the variation in the results are due to the quality of the learned embeddings and not to the implementation of the downstream task models -- while more sophisticated options are detailed in Section~\ref{conclusion_section}, our benchmarks show that aligned embeddings are indeed an extremely promising area of exploration.

Finally, it is important to stress that given the novelty of the setting (as discussed in Section~\ref{usecasesection}) and the differences with cross-space tasks in NLP settings, \textit{prima facie} plausible baselines are actually not good candidates for the scenarios at hand. For example, even in the presence of high-quality cross-shop tracking, joint embeddings cannot be trained on cross-shop sessions due to data sparsity; as another example, recent alignment techniques that are successful for word spaces (e.g.~\cite{Bianch2019}) rely on the assumption that either many labeled pairs are available, or that the vast majority of the embedding space is comprised by pairs of identical items; other interesting ideas, such as using product titles for a similarity metrics, would require uniformity in meta-data, which is an assumption that no proposed models make. Framed as a zero-shot inference,~\textit{multi-shop} predictions are a relatively new challenge and we hope our work (and dataset) to be a long-lasting contribution to the community.

\subsection{Next Event Prediction across shops}
For the cross-shop prediction task, we sampled 12510 browsing sessions over a month (\textit{not included in the training set}) for distinct users that visited \textbf{Shop A} and \textbf{Shop B} within the same day.

\subsubsection{Quantitative evaluation}
We benchmark the cross-shop methods from Section \ref{mapping_section} against three baselines of increasing sophistication:

\begin{itemize}
    \item \textit{popularity model} (PM): while trivial to implement, leveraging product popularity is by far the most common heuristic in the industry for the zero-shot scenario, and it has been proven to be surprisingly competitive in many e-commerce settings against statistical and neural approaches~\cite{Dacrema:2019}; also, given that popular products are more likely to be on display and generate a classic ``rich get richer dynamics'', quantitative results for \textit{PM} are likely to overestimate its efficacy and therefore raising the bar for other methods;
    \item \textit{activity-based model} (AM): a \textit{semi-supervised} model, inspired by evidence from NLP literature in which \textit{some} supervision goes a long way in helping with the alignment process~\cite{artetxe-etal-2017-learning}; in particular, the model  leverages domain knowledge (sport activity for each product) that is however not directly related to the mapping we are trying to learn. We randomly sample 20 products from \textbf{Shop A} of category $S$ and from \textbf{Shop B} within the same category, using activities as "known similar regions", and we then we learn a mapping function using standard linear regression from the centroid of the sampled products from the two spaces;
    \item \textit{iterative alignment model} (NM): state-of-the-art unsupervised method from~\cite{Artetxe18}, originated in the NLP literature: the model is quite sophisticated and its performances in this scenario shed interesting insights on how peculiar the task of aligning product embeddings is (as compared to word embeddings); in a nutshell,~\textit{NM} leverages the structure of embedding spaces to build an initial weak dictionary; the dictionary is then used to bootstrap a self-learning process, which iterates through mapping and dictionary optimization, until convergence is reached.
\end{itemize}

\begin{table}
  \caption{NDCG@10 for supervised and unsupervised models in the \textit{First Item Prediction} (FIP) and \textit{Any Item Prediction} (AIP) tasks: best results per type are highlighted in bold.}
  \label{tab:results_first}
  \begin{tabular}{lccc}
    \toprule
    Model & Type & FIP & AIP\\
    \midrule
    PM & Unsupervised & 0.00232 & 0.00297\\
    NM & Unsupervised & 0.00097 & 0.00112\\
    IM & Unsupervised & \textbf{0.01506} & \textbf{0.01628}\\
    \hline
    AM & Semi-supervised & 0.00108 & 0.00121\\
    UM & Supervised & 0.02741 & 0.02854\\
    TM & Supervised & \textbf{0.03786} & \textbf{0.04501}\\
    \bottomrule
\end{tabular}
\end{table}

Table \ref{tab:results_first} reports \textbf{NDCG@10} for all models for two prediction tasks: \textit{First Item Prediction} (FIP) and \textit{Any Item Prediction} (AIP). \textit{FIP} is the ability of the proposed model to guess the first product in the target shop, while \textit{AIP} is the ability to guess \textit{any} product found in the session in the target shop. Unsurprisingly, fully supervised models outperform all other methods; among unsupervised models, the \textit{IM} model we propose is the best one, resulting in a 549\% increase over the industry baseline and even significantly \textit{beating the semi-supervised baseline} \textit{AM}\footnote{Generally speaking,~\textit{AM} seems to overfit on common categories and turns out to be worse than the simple \textit{PM} model.}; the performance gap between \textit{IM} and \textit{NM} highlights that straightforward implementation of SOTA models from NLP does not guarantee the same results in the product scenario. Among supervised models, \textit{TM} outperforms \textit{UM} on \textit{FIP} and provides a 1530\% increase over the industry baseline; to test if \textit{TM} improves significantly with data quantity, we ran an additional test on a separate cross-shop dataset from our network of clients: \textit{TM} results on this second set for \textit{FIP}/\textit{AIP} are $0.066$/$0.071$, and $0.021$/$0.023$ for \textit{UM}, showing that indeed the seq2seq architecture may be the best option for use cases in which significant amount of cross-shop behavior has been tracked already.

In the spirit of ablation studies, we generated predictions on the same \textit{cross-shop} dataset using \textit{IM} but employing instead \textit{low-scoring} embedding spaces, to assess whether picking optimized vs non-optimized spaces make a difference in the zero-shot prediction task: the reported \textbf{NDCG@10} for this setting is $0.005$, which is \textit{significantly lower} than the reported best score obtained with the optimized embeddings.

\subsubsection{Qualitative evaluation}
Given the novelty of the experimental settings, a qualitative evaluation is important as well to interpret the outcome of the benchmarks above: is the alignment of the two spaces capturing important human-level concepts? We devised two additional tests to answer these questions. First, we test the aligned embeddings in a ``cross-shop activity prediction'' task: using the same setup from Section~\ref{qualitative_val}, we train an MLP on \textbf{Shop A} \textit{aligned} embeddings and use it without additional training on \textbf{Shop B} \textit{aligned} embeddings. The mean accuracy for activity prediction over 5 runs is $\mu=0.73$ ($SD=0.002$), confirming that the alignment process can effectively transfer learning from \textbf{A} to \textbf{B}. 

\begin{figure}
  \centering
  \includegraphics[width=\linewidth]{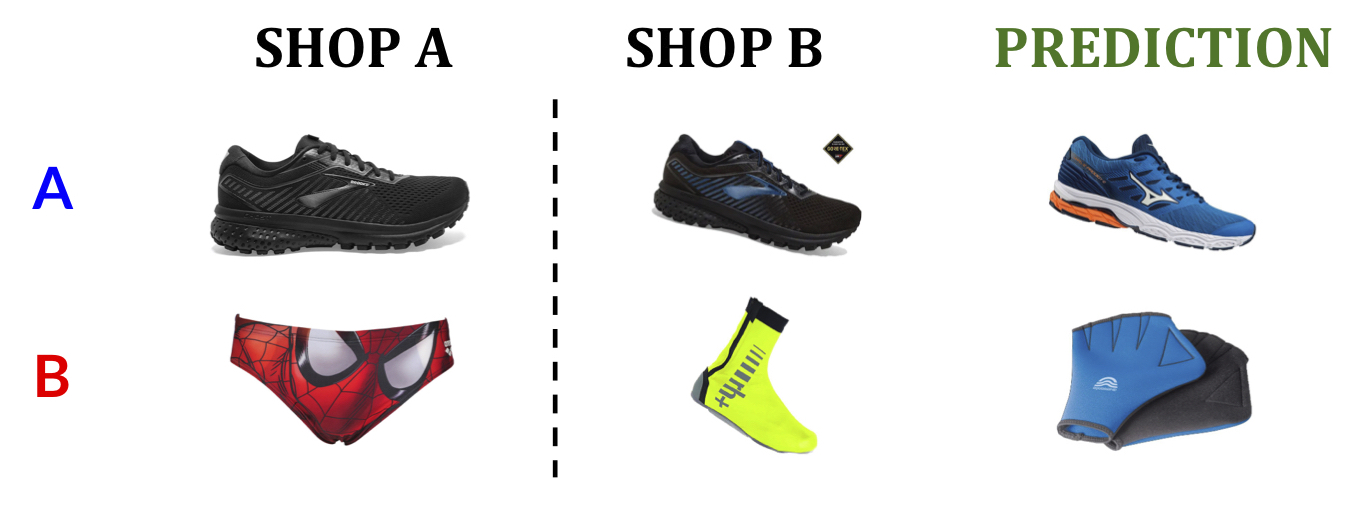}
  \caption{Two sample sessions from the \textit{cross-shop} portion of the dataset: \textbf{Session A} is a session with stable shopping intent (i.e. "running") and model prediction is wrong but plausible; \textbf{Session B} is made of two disconnected intents and model prediction is significantly wrong.}
  \label{error_examples}
  \Description{Pipeline functional overview.}
\end{figure}

Second, we perform \textit{error analysis} on several misclassified cases. Our exploration highlights that pure quantitative measures - such as \textbf{NDCG@10} - are great at capturing high-level patterns of efficacy for the chosen models, but cannot capture important differences in particular cases of \textit{cross-shop} predictions. If we think about the particular task of zero-shot recommendation, \textbf{NDCG@K} is asking the model to pick \textit{the} one correct product out of \textit{several thousands}, which is likely to underestimate the practical efficacy of the proposed recommendations. Instead of just computing an hit/miss ratio for \textbf{NDCG@K}, we ran the \textit{IM} model on the test set recording, for every "miss", the distance in the shared embedding space between the target product and the predicted one; we then order these wrong predictions according to the magnitude of the error, and analyze sessions from the top and bottom of the distribution. Interestingly enough, sessions with a small recorded error are the ones that looks coherent to a human observer, as in \textbf{Session A} in Figure~\ref{error_examples}, where running shoes from \textit{Brooks} manufacturer are confused by the model with running shoes from \textit{Mizuno} manufacturer; when error margin gets big, situations like \textbf{Session B} are more common: products in the same \textit{cross-shop} session are very different, since the shopper intent may have drifted between the two visits - the prediction of the model is significantly off (wrong object, wrong manufacturer, wrong sport activity). To try and quantify the proportion of ``reasonable'' mistakes, we train an MLP mapping the target and the predicted product to a sport activity (as in Section \ref{qualitative_val}), and comparing the first predicted activity versus the ground truth: this  model achieves \textit{zero-shot} accuracy of $0.44$, which raises to $0.66$ if we consider just sessions whose error distance is below the median (i.e. sessions with more "stable" intent). 

All combined, these findings suggest that models are successfully transferring shopping intent and they are likely to perform well in practice for all the sessions in which intent across shop is consistent, even when the predicted item is not \textit{exactly} a match (e.g. \textbf{Session A} in Figure \ref{error_examples}; cases like \textbf{Session B} are unlikely to be solvable anyway).

\begin{figure}
  \centering
  \includegraphics[width=\linewidth]{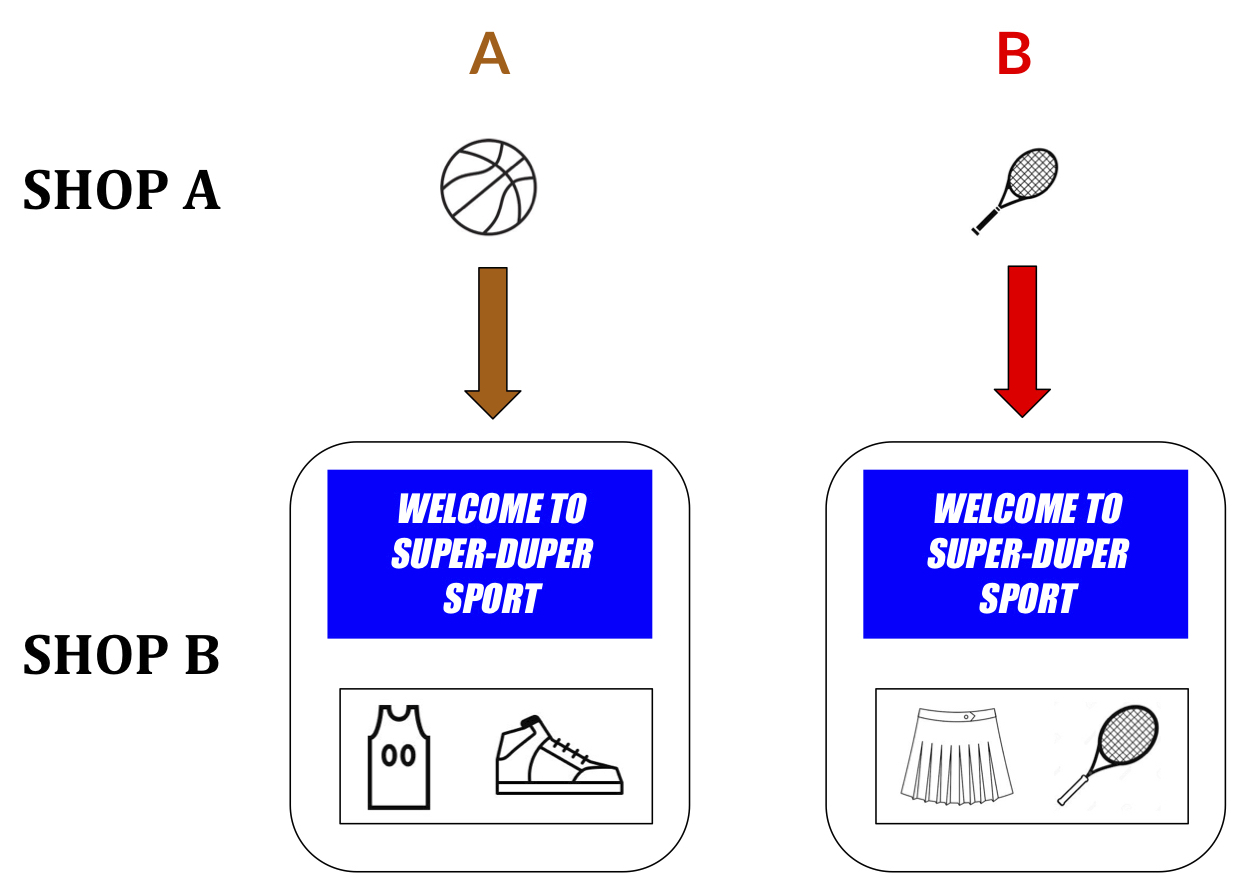}
  \caption{Landing pages can be customized in real-time by transferring intent from previous shops to the current one: by focusing on the general activity, instead than the exact product, we make the task easier for the model and unlock more use cases for the clients. In this example, \textbf{Shop B} presents a basketball-themed page to \textbf{User A} and a tennis-themed page to \textbf{User B}.}
  \label{landing_page}
  \Description{Pipeline functional overview.}
\end{figure}

\subsection{Personalized Type-Ahead across shops}
\begin{figure}
  \centering
  \includegraphics[width=7.5cm]{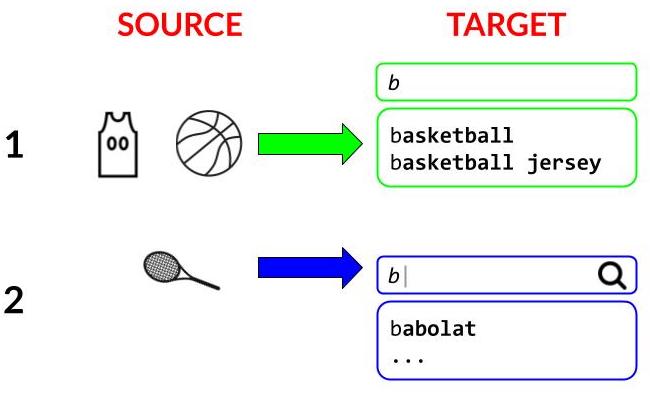}
  \caption{Two sessions illustrating cross-shop personalization for type-ahead suggestions: the same prefix ``b'' on the target website triggers different completion depending on intent transferred from the source shop.}
  \label{sigir_cross}
  \Description{Example of cross-shop personalization for type-ahead suggestions.}
\end{figure}

As a second, less direct application of aligned embedding spaces, we propose to exploit product embeddings in a conditional language model, to provide personalized type-ahead suggestion to incoming users on a target shop (Fig.~\ref{pred_pic}). We deploy the same type-ahead framework we proposed in~\cite{CoveoECNLP20}, in which an encoder-decoder architecture is employed to first encode user intent, and then use an LSTM-powered char-based language model to sort query completions by their probability (please refer to the paper for architectural details): as illustrated by Fig.~\ref{sigir_cross}, if the user's session is basketball-themed (\textbf{1}), we expect completions like \textit{basketball jersey} for prefix \textit{b}; if it is tennis-themed (\textbf{2}), the same prefix may instead trigger a tennis brand like \textit{babolat}.

\subsubsection{Quantitative evaluation}
Table~\ref{tab:results_cross} shows the results of our quantitative benchmarks for the \textit{cross-shop} scenario, comparing a non-personalized baseline to models performing transfer learning. For the personalized predictions, we train a conditional language models on the target shop first. At prediction time, we feed to the target shop model the \textit{aligned} embeddings from the source shop, perform average pooling in the encoder~\cite{CoveoECNLP20}, and read off the decoder conditional probabilities of the target query suggestions.

We use ~\textit{Mean reciprocal rank} (\textbf{MRR}) as our main metric, as a standard in the auto-completion literature: \textbf{MRR@k} is \textbf{MRR} measured by retrieving from the model the first $k$ suggestions. In our experiments, $k$ is set to $5$ to mimic the target production environment:

\begin{equation}
\mathrm{MRR}=\frac{1}{|Q|} \sum_{i=1}^{|Q|} \frac{1}{\operatorname{rank}_{i}}
\label{eq:mrr}
\end{equation}
where ${\operatorname{rank}_{i}}$ is the position of the first relevant result in the ${i}$-th query and ${Q}$ is the total number of queries.

The best supervised models provide up to 600\% uplift, but even the \textit{purely unsupervised} model significantly outperforms the non-personalized model, establishing that transferring intent is \textit{significantly} better than treating all incoming shoppers as \textit{new}; for mid-size and large retailers, capturing the interest of even a small percentage of these users may provide significant business benefits.

\begin{table}
  \caption{MRR@5 in the \textit{cross-shop} scenario, for different seed length (SL), for shoppers going from \textbf{A} to \textbf{B} and issuing a query there.}
  \label{tab:results_cross}
  \begin{tabular}{lcc}
    \toprule
    Model & SL=0 & SL=1\\
    \midrule
    PM & 0.001 & 0.045\\ 
    Vec2Seq+IM & 0.005 & 0.050\\
    Vec2Seq+UM & 0.003 & 0.055\\
    Vec2Seq+TM & \textbf{0.007} & \textbf{0.062}\\
  \bottomrule
\end{tabular}
\end{table}

\subsubsection{Qualitative evaluation}
\begin{figure}
  \centering
  \includegraphics[width=7.5cm]{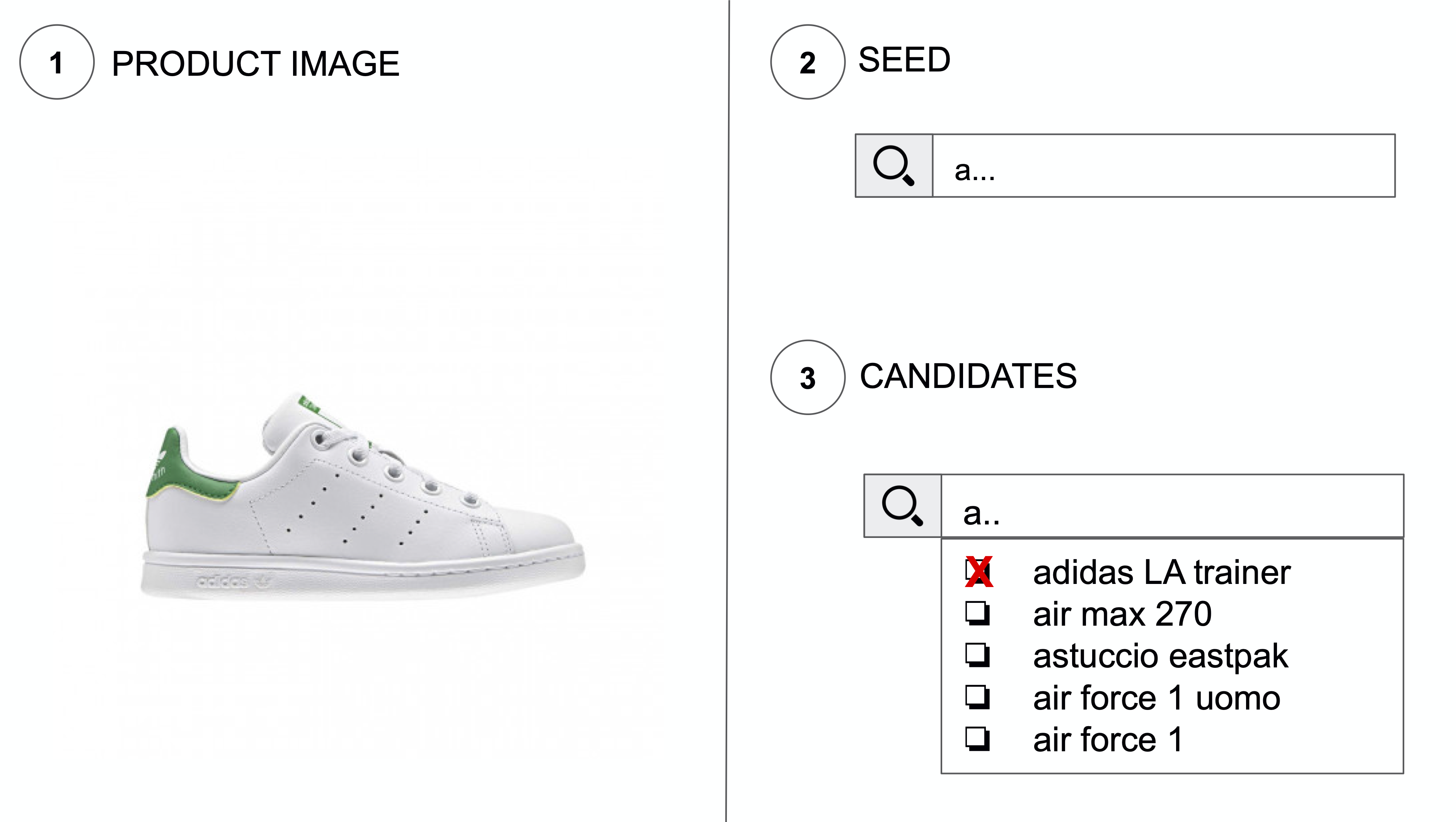}
  \caption{Example of a stimulus in the qualitative user study.}
  \label{qual_validation_pic}
  \Description{Example of the relevance task for qualitative evaluation.}
\end{figure}

Quantitative benchmarks provide empirical evidence on the overall efficiency of personalization, but as discussed,~\textit{cross-shop} sessions ``in the wild'' sometime show drifting intent across sites. To specifically test how much the transferred intent is able to capture \textit{semantic similarity} across the two aligned spaces, we devise a small user study. We recruited 20 native speakers, whose age ranged between 22 and 45; subjects (Figure \ref{qual_validation_pic}) were presented with a product image from S-Shop (1), a seed character (2) and were asked to pick the most relevant completion among 5 candidates (3). The <\textit{product image}, \textit{seed}> pairs are taken from representative queries from the \textit{cross-shop} set, for a total of 30 stimuli for each subject; five candidate queries are chosen by first retrieving the top 35 candidates from the unconditioned model, and then sampling without replacement. By collecting semantic judgment directly, our prediction is that the performance gain from personalization will be higher, since the study should eliminate the popularity bias implicit in search logs.

\textit{PM}, \textit{IM} and \textit{TM} are tested against the collected dataset, resulting in a \textbf{MRR@5} of, respectively, $0.076$, $0.123$ and \textbf{$0.138$}; \textit{TM} accuracy with $SL=1$ is $81\%$ higher than \textit{PM}, supporting our hypothesis that the aligned embeddings successfully transfer user intent in the zero-shot scenario.

\section{(Vector) Space, the final frontier: what's next?}
\label{conclusion_section}
In \textit{this} work we detailed a machine learning pipeline for behavioral data ingestion finalized to train \textit{prod2vec} models, i.e. generate neural product representations for several downstream prediction tasks. In the first part, we focused on training the best embeddings as judged by quantitative and qualitative validation. Product representations have been found to be increasingly useful in many e-commerce scenarios~\cite{tagliabue-etal-2020-grow}, but the understanding of them in realistic industry scenarios is still incomplete; on this point, it is telling that several findings of a recent hyperparameter study (\cite{Caselles18}) could \textit{not} be replicated in our context. For this reason, we believe that the \textit{within-shop} training portion of our pipeline can provide a useful assessment for production systems in the industry, starting from our validation best practices and engineering considerations. Prompted by the industry need for few-shots and scalable personalization and practical deployment concern of our growing client network, the second part of \textit{this} work was focused on generalizing product spaces to address the \textit{cross-shop} scenario depicted in Fig.~\ref{pred_pic}. We devised and tested several models with varying degrees of supervisions, and, again, supplemented our quantitative benchmarks with additional qualitative tasks to gain a better understanding of model performances in this new scenario. All in all, the evidence provided is a strong argument in favor of our initial research hypothesis, i.e. that embedding spaces from two shops can be successfully aligned, so that zero-shot predictions can be performed in a principled way.

While the theoretical and engineering foundations of the platform have proven to be solid and crucial in solving retail problems at scale, our roadmap is focused on taking these ideas even further. Broadly speaking, we can classify open issues in two categories, \textit{research} and \textit{product} improvements:

\begin{itemize}
    \item \textbf{research}: since i) there is independent demand for general purpose \textit{prod2vec} models, ii) universal tracking is still available in a limited fashion, we did not test end-to-end learning by using \textit{cross-shop} predictions as the optimization task \textit{directly}; as more data becomes available, it is a natural extension to the methods proposed in \textit{this} work. Moreover, as highlighted in Section~\ref{experiments}, significant optimization can be made to neural architectures for downstream tasks now that \textit{this} study first established the viability of aligned embeddings to capture user's intent across shops;
    \item \textbf{product}: as discussed in Section~\ref{usecasesection}, online retailers are facing increasing pressure to deliver relevant experiences to incoming customers; the question is not \textit{whether} personalization should be done, but \textit{how soon} into the shopper journey it can be done. We are actively working with several fashion groups to deploy cross-shop models and perform live A/B testing of the proposed methods; in our growing SaaS network of retailers, we believe more and more global multi-shop opportunities will soon benefit at scale from our research. 
\end{itemize}

On a final note, we hope that curating the first dataset of its kind will help drawing increasing attention from industry and academic practitioners to these important business scenarios. SaaS providers with an extensive network of clients are ideally suited to leverage transfer learning techniques, including the alignment of embeddings here introduced; at the same time, some of the biggest traditional retailers in the world are indeed \textit{multi-brand} groups, and they could ``transfer knowledge'' between their brands to provide personalization in an hyper-competitive, data-driven market.

In a time characterized by growing concerns on long-term storage of personal data \cite{Voigt:2017:EGD:3152676}, we \textit{do} believe that small-data learning will be a distinctive feature for successful players in this space. 

\begin{acks}
Thanks to the anonymous reviewers and Piero Molino for comments on previous versions of \textit{this} work. Thanks to Caterina \textit{Caterí} Vernieri, who brought LaTeX magic and stellar T-factor into our paper and our lives (not in this order). Special thanks to Andrea Polonioli for his support.
\end{acks}

\bibliographystyle{ACM-Reference-Format}
\bibliography{sigir_refs}

\appendix
\section{DATA PIPELINE WITH PAAS SERVICES}
\label{paasdetailssection}
For practitioners in the same industry, Figure~\ref{pic:etl_aws} gives a high-level sketch of how the chosen PaaS services fit together in the pipeline:

\begin{figure}[h]
    \includegraphics[width=8.5cm]{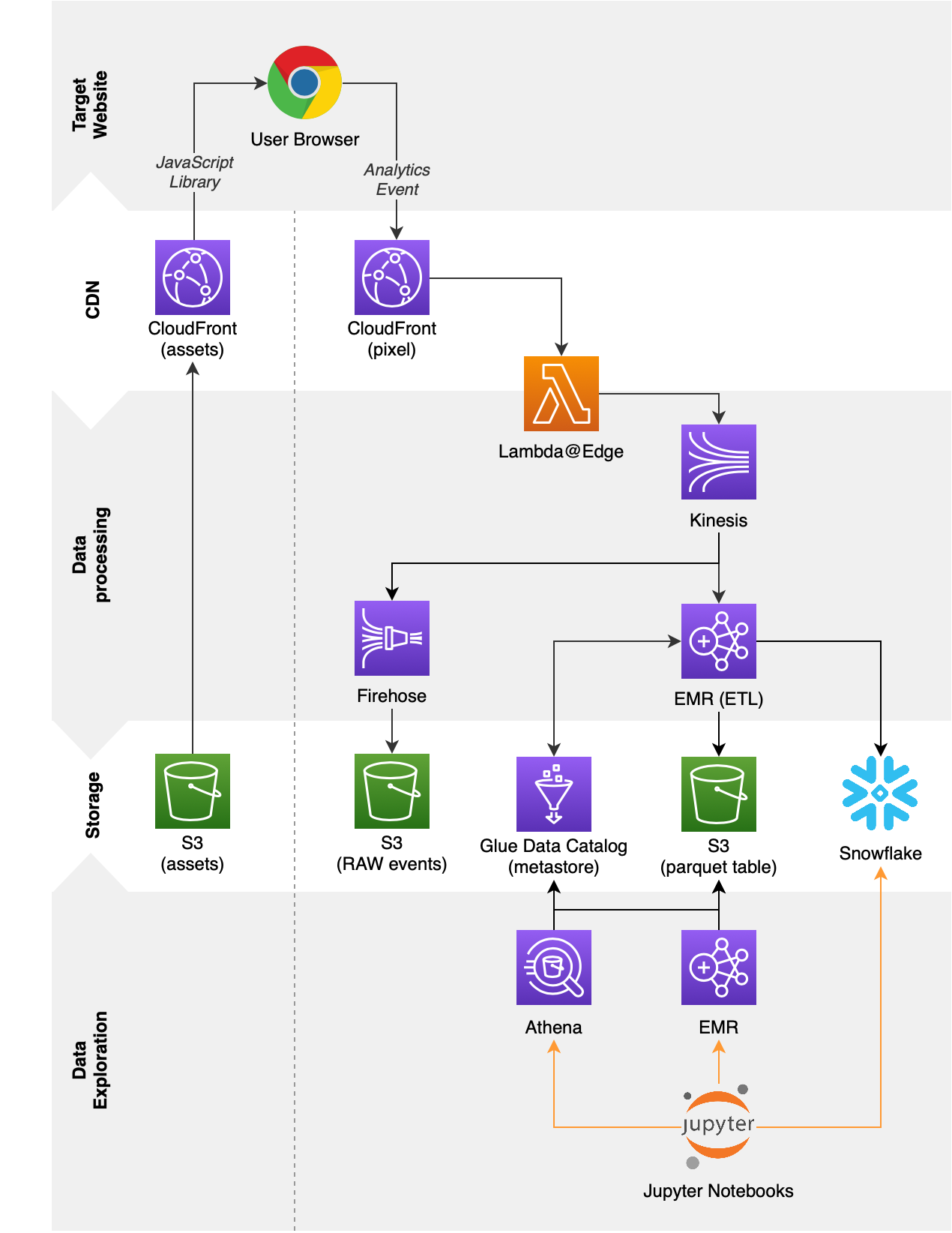}
    \caption{Cloud-based data ingestion pipeline.}
    \label{pic:etl_aws}
\end{figure} 

\begin{itemize}
    \item the Javascript library is stored on S3 and globally distributed through AWS CloudFront\footnote{\url{https://aws.amazon.com/cloudfront/}};
    \item the pixel endpoint is reachable through AWS CloudFront, to ensure high performances;
    \item incoming events are processed by an AWS Lambda@Edge\footnote{\url{https://aws.amazon.com/lambda/edge/}} and streamed to internal consumers by using AWS Kinesis\footnote{\url{https://aws.amazon.com/kinesis/}};
    \item AWS Firehose\footnote{\url{https://aws.amazon.com/kinesis/data-firehose/}} is used to persist all the RAW events in S3\footnote{\url{https://aws.amazon.com/s3/}} for future re-processing;
    \item the ETL processing is done in an AWS EMR\footnote{\url{https://aws.amazon.com/emr/}} Cluster; normalized and sessionized events are then stored on S3 in a Parquet format;
    \item tables metadata are stored in AWS Glue Data Catalog\footnote{\url{https://aws.amazon.com/glue/}}; data are made querable with Spark-SQL on EMR and AWS Athena\footnote{\url{https://aws.amazon.com/athena/}}.
    \item data are also stored in Snowflake\footnote{\url{https://www.snowflake.com/}} as part of our project for a future simplification of our data warehouse practices.
\end{itemize}

\end{document}